\documentclass[usenatbib]{mn2e}
\usepackage{times}
\usepackage{graphicx}
\def\mnras{MNRAS}
\def\apj{ApJ}
\def\apjl{ApJ Letters}

\def\he3{$^3$He~{\small II}}

\pubyear{2008}

\author[Bagla and Loeb]{J.S.Bagla$^1$ and Abraham Loeb$^2$ \\
  $^1$ Harish-Chandra Research Institute,  Chhatnag Road, Jhusi,
  Allahabad 211019, India. \\
  $^2$ Harvard-Smithsonian Center for Astrophysics, Mail Stop 51, 60 Garden
  Street, Cambridge, MA 02138, USA \\
  E-Mail: $^1$ jasjeet@hri.res.in, $^2$ aloeb@cfa.harvard.edu
}

\title[Probing the IGM with \he3]{The hyperfine
  transition of $^3$He~{\Large II} as a probe of the
  intergalactic medium}  

\label{firstpage}

\def\LaTeX{L\kern-.36em\raise.3ex\hbox{a}\kern-.15em
    T\kern-.1667em\lower.7ex\hbox{E}\kern-.125emX}

\begin{document}

\maketitle

\begin{abstract}
We explore the prospects of using the hyperfine transition of \he3 as
a probe of the intergalactic medium. The emission signal from ionized
regions during reionization is expected to be anti-correlated with
$21$~cm maps. The predicted emission signal from Ly-$\alpha$
blobs at lower redshifts is detectable with future radio
observatories.

\end {abstract}




\section{Introduction}

Galaxies form when gas in highly overdense haloes cools and collapses
to form stars \citep{1953ApJ...118..513H, 1977MNRAS.179..541R,
1977ApJ...211..638S, 1977ApJ...215..483B, 1991ApJ...379...52W}.  The
formation of the first stars \citep{2009arXiv0905.0929B,
2007ARA&A..45..565M, 2007ARA&A..45..481Z, 2004ARA&A..42...79B} led to
emission of UV radiation that re-ionized the intergalactic medium (IGM)
of hydrogen and helium \citep{2001ARA&A..39...19L}.

A number of methods can be used to probe the reionization process
\citep{2006ARA&A..44..415F}, but the most promising probe appears to
be the redshifted emission and absorption due to the hyperfine
transition of neutral hydrogen with a rest frame frequency of
$1.42$~GHz (for a recent review see \citet{2006PhR...433..181F}).

Singly ionized $^3$He has a hyperfine transition similar to the $21$~cm
spin-flip transition of hydrogen.  The astrophysical relevance of this
transition was first pointed out by \citet{1957IAUS....4...92T}.
Early studies focused on observing the transition from the
interstellar medium, specifically from emission line nebulae
\citep{1966AZh....43.1237S, 1967ApJ...149...15G, 1967PhRvL..18..433G}.
Several papers \citep{1967SvA....11..233D, 1975MNRAS.171..375S,
2006PhR...433..181F} pointed out that this transition may potentially
be used to probe early stages of galaxy formation.

Following the early theoretical work, attempts were made to observe
this signal from nearby H~{\small II} regions
\citep{1971ApJ...168L.125P}, leading to the first detection by
\citet{1979ApJ...227L..97R}.  This line has been detected from a
number of H~{\small II} regions as well as a few planetary nebulae
\citep{2000ApJ...531..820B, 2006ApJ...640..360B, 2007Sci...317.1171B}.
The primordial abundance of $^3$He can be used to constrain big bang
nucleosynthesis \citep{2007ARNPS..57..463S}.  The main challenge in
observing this transition lies in separating it from recombination
lines of hydrogen, which can be almost as strong if not stronger.

In this paper we explore the potential use of the hyperfine transition
of \he3 for studying the epoch of reionization (EoR) and the
post-reionization IGM.  In \S 2 we discuss aspects of the hyperfine
transition for \he3 and compare it with the corresponding transition
of neutral hydrogen.  Some sources and absorbers of this transition in
the IGM are enumerated in \S{3}.  Finally, we consider in \S{4} the
most promising sources from where \he3 emission may be observed, and
how future observations may constrain the EoR and the IGM. Throughout
the paper, we adopt the WMAP-5yr values for the cosmological
parameters \citep{2008arXiv0803.0547K}.

\section{Hyperfine Transition of \he3}

We begin with a comparison of hyperfine transitions of H I and {\he3}.
For {\he3}, the frequency of hyperfine transition is $8.66$~GHz, much
higher than $1.42$~GHz frequency of the $21$~cm transition of hydrogen.
A higher frequency for the \he3 hyperfine transition implies that the
emission signal from the EoR is at frequencies where foregrounds are
not as problematic as they are for $21$~cm observations.

The magnetic moment of the $^3$He nucleus is negative, and so the
singlet is the excited state in the hyperfine transition while the
triplet is the ground state.  This is opposite of the situation in
neutral hydrogen.  As a result of the ground state being the triplet,
emission is suppressed.  The Einstein coefficient for the spontaneous
decay of this transition is $A_{10} = 1.96 \times 10^{-12}$s$^{-1}$
\citep{1966AZh....43.1237S, 1967ApJ...149...15G},
which is $\sim 700$ times larger than the corresponding coefficient
for the hyperfine transition of hydrogen.
The effective gain is reduced by a factor of $3$ due to the excited
state being the singlet.  The higher rate of emission implies a
shorter lifetime for electrons in the excited state and this has
interesting implications for observables, as shown in the discussion
below.

The number density of $^3$He atoms is a small fraction of the hydrogen
density, the primordial ratio from big bang nucleosynthesis being
close to $1.1 \times 10^{-5}$ \citep{2007ARNPS..57..463S}.  The
ionization potential for hydrogen is $13.6$eV, while the ionization
potential for helium is $24.6$eV for the first electron and $54.4$eV
for the second electron.

In analogy with the $21$~cm transition, the optical depth for absorption
of radiation from a background source by \he3 is given by,
\begin{eqnarray}
\tau_\nu &=& \frac{1}{32\pi}\frac{h c^3 A_{10}}{k_B T_s \nu_0^2} \frac{x_{He
  II}   n_{^3He}} {(1+z) (dv_\parallel/dr_\parallel)} \nonumber \\
&=&  6 \times 10^{-7} ~~~~ \frac{x_{He II}}{T_s/{\rm K}} (1+\delta) (1+z)^{3/2}
 \nonumber \\
&& ~~~~~~~~~ \times ~~~~~~
\left(\frac{n_{^3He}/n_{H}}{1.1 \times 10^{-5}}
  \right)\left[\frac{H(z)}{(1+z) (dv_\parallel/dr_\parallel)}\right] .
\end{eqnarray}
where $n_{\rm H}\propto (1+z)^3$ is the primordial hydrogen density as
a function of redshift $z$,
$H(z)=H_0[\Omega_\Lambda+\Omega_m(1+z)^3]^{1/2}$ is the Hubble
expansion rate (with $\Omega_\Lambda$ and $\Omega_m$ being the
present-day density parameters of the cosmological constant and
matter, $dv_\parallel/dr_\parallel$ is the comoving radial derivative
of the radial velocity component of the cosmic gas along the
line-of-sight (which equals $H(z)/(1+z)$ for the uniform background
Universe), $\delta$ is the local overdensity of baryons, and $x_{He
II}$ is the fraction of Helium atoms in the singly ionized state.  The
spin (excitation) temperature, $T_s$, is defined through the ratio
between the number densities of hydrogen atoms in the excited and
ground state levels,
\begin{equation}
{n_1\over n_0}={g_1\over g_0}\exp\left\{-{T_\star\over T_s}\right\},
\label{eq:spin}
\end{equation}
where subscripts $1$ and $0$ correspond to the excited and ground
state levels of the \he3 transition, $T_\star=h\nu/k_B=0.42$K is the
temperature corresponding to the transition energy, and
$(g_1/g_0)=1/3$ is the ratio of the spin degeneracy factors of the
levels.

The primary process that couples $T_s$ with the gas temperature
$T_{\rm gas}$ is the scattering of free electrons and \he3 ions.
There are two time scales of interest: the lifetime of an electron in
the excited state, and the typical time between electron-ion
collisions.  Furthermore, we have to take into account the efficacy of
collisions in setting the spin temperature.  The ratio of the
collision rate to the lifetime scales as $(1+z)^3 (1 + \delta)$.
Thus, $T_s$ and $T_{\rm gas}$ couple well at high redshifts or in
regions with high overdensity.  The Wouthuysen-Field
effect \citep{field1958, 1959ApJ...129..536F, 1952Phy....18...75W} is
less relevant in coupling $T_s$ with the gas temperature $T_{\rm gas}$
due to the proximity of the Ly-$\alpha$ lines for the two isotopes of
Helium \citep{2006ApJ...651....1C}.  The color temperature is generally
negative due to this proximity.  
In some situations in the interstellar medium, the proximity of the Ly-$\alpha$
lines for the two isotopes and an O~{\small III} line can even lead to
negative spin temperature \citep{1985ApJ...290..578D} in low density
environments where collisions are not relevant and significant amount of
He~{\small III} is present. 

For a fixed ionization fraction and a homogeneous Universe, the time
evolution of the density of \he3 ions in the ground state is given by,
\begin{eqnarray}
\left( \partial_t  +   3{{\dot a}\over a} \right)
 n_0  &=& -n_0\left(C_{01}+B_{01}I_\nu\right) \nonumber \\
 && ~~~ +  n_1  \left(C_{10}+A_{10}+B_{10}I_\nu\right),
\label{eq:evolution}
\end{eqnarray}
where $a(t)=(1+z)^{-1}$ is the cosmic scale factor, the $A$'s and
$B$'s are the Einstein rate coefficients, the $C$'s are the
collisional rate coefficients, and $I_\nu$ is the blackbody intensity
in the Rayleigh-Jeans tail of the Cosmic Microwave Background (CMB),
namely $I_\nu=2k_BT_{cmb}/\lambda^2$ with the transition wavelength
$\lambda=c/\nu=3.46$~cm.  The $0\rightarrow 1$ transition rates can be
related to the $1\rightarrow 0$ transition rates by the requirement
that in thermal equilibrium with $T_s=T_{cmb}=T_{\rm gas}$ (where
$T_{\rm gas}$ is the gas kinetic temperature), the right-hand-side of
Eq. (\ref{eq:evolution}) should vanish with the collisional terms
balancing each other separately from the radiative terms.  This
implies that $(C_{01}/C_{10})=(g_1/g_0)\exp\{-T_\star/T_{\rm gas}\}$.
We also use the standard relations $B_{10}=(\lambda^3/2hc) A_{10}$ and
$B_{01}=(g_1/g_0)B_{10}$ \citep{1986rpa..book.....R}.  The collisional
rates are proportional to the electron density $n_e$ and depend on
$T_{\rm gas}$.

Equation (\ref{eq:evolution}) can
be simplified to the form,
\begin{eqnarray}
{d\Upsilon \over dz} & = & -\left[H(1+z)\right]^{-1}
\left[-\Upsilon(C_{01}+B_{01}I_\nu) \right. \nonumber \\
&& \left. +
(1-\Upsilon)(C_{10}+A_{10}  + B_{10}I_\nu)\right],
\label{eq:upsilon}
\end{eqnarray}
where $\Upsilon\equiv n_0/(n_0+n_1)$. In the redshift range of
interest, the Hubble expansion rate can be ignored relative to the
atomic rate coefficients and so the left-hand-side of equation
(\ref{eq:upsilon}) can be approximately set to zero. For $T_s\gg
T_\star$ and $T_{\rm gas}\gg T_\star$, this yields the simple result
\citep{2006PhR...433..181F,2008PhRvD..78j3511P},
\begin{equation}
T_s^{-1}= {T_{cmb}^{-1} +x_cT_{\rm gas}^{-1} \over 1 + x_c} ,
\label{simple}
\end{equation}
where $x_c=(T_\star/T_{\rm cmb})(C_{10}/A_{10})$.

\begin{figure}
\begin{center}
\includegraphics[width=3truein]{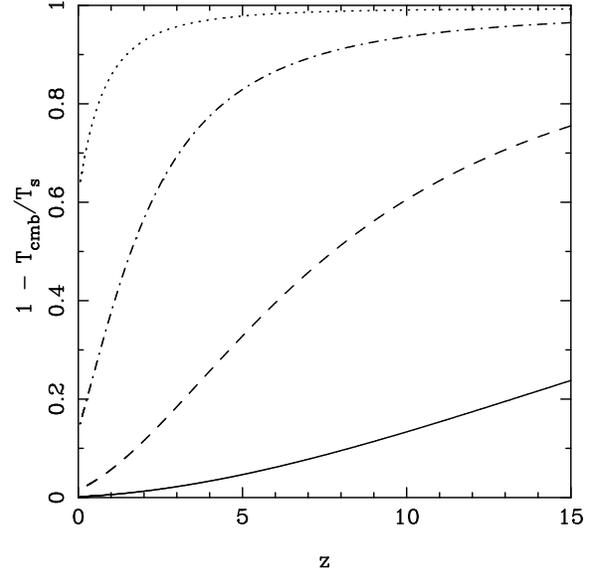}
\end{center}
\caption{The coefficient $(1 - T_{cmb}/T_s)$ (to which the brightness
temperature in Eq. (\ref{Tb}) is proportional) as a function of
redshift for regions of different values of overdensity, $\delta$.
The spin de-excitation rate is expressed as a fraction $f_c$ of the
value derived by \citet{1967ApJ...149...15G}.  Different lines
correspond to different values of $f_c (1+\delta)$, namely $1$ (solid
line), $10$ (dashed line), $10^2$ (dot-dashed line), and $10^3$
(dotted line).  We assume that hydrogen is fully ionized and helium is
singly ionized, and that the gas temperature is $10^4$~K.}
\end{figure}

The charged \he3 ions interact with electrons through Coulomb
collisions. The standard collision rate in a plasma receives equal
contributions per logarithm of the impact parameter between the
electron and the \he3 ion, and so it includes the cumulative effect of
many small-angle deflections of the electron
\citep{1998ppim.book.....S}.  However, most of the small angle
scatterings will not result in a spin-flip transition for the bound
electron.  The maximum impact parameter of interest for exciting the
\he3 transition is set by the requirement that the energy transfer per
single electron-ion collision be larger than the \he3 transition
energy
\footnote{Following \citet{field1958}, we ignore coupling of the
incoming electron with the spin of the nucleus due to the larger mass
of the nucleus.}.  For the distance of closest approach of a thermal
electron relative to a \he3 ion, the typical energy transfer is $\sim
(m_e/m_{\rm 3He}) k_B T_{\rm gas}$, amounting to $\sim 1.7$~K for
$T_{\rm gas}=10^4$~K.  The transition energy of $0.42$K implies in
this case a Coulomb logarithm of $\sim \ln(1.7/0.42)=1.4$, which is an
order of magnitude smaller than its maximum value, obtained from the
cumulative effect of all scatterings with a smaller energy
exchange\footnote{http://wwwppd.nrl.navy.mil/nrlformulary/NRL$\_$FORMULARY$\_$09.pdf}.
This implies that only a small subset ($< 0.1$) of all electron-ion
collisions contribute.  An estimate for the collisional coupling
efficiency was derived by \citet{1967ApJ...149...15G}, based on the
approach pioneered by \citet{1956ApJ...124..542P, field1958,
1959ApJ...129..536F}.  We quantify our collisional rate relative to
this value.

Figure~1 shows the quantity $(1 - T_{cmb}/T_s)$ as a function of
redshift for regions with different gas density contrast $(1+\delta)$
relative to the cosmic mean.  The results were derived assuming that
hydrogen is fully ionized and helium is singly ionized, and that the
gas temperature is $10^4$~K.  The spin de-excitation rate is
proportional to the collisional rate, which in turn is is proportional
to $(1 + \delta)$.  We take the spin de-excitation rate to be a
fraction $f_c$ of the \citet{1967ApJ...149...15G} rate.  Different
lines correspond to different values of $f_c (1+\delta)$, namely $1$
(solid line), $10$ (dashed line), $10^2$ (dot-dashed line), and $10^3$
(dotted line).  The thick solid line that runs along the curve for
$f_c (1+\delta)=1$ shows the collisional coupling estimate of
\citet{1967ApJ...149...15G}.  A more detailed calculation of the
collisional rate can be made using spin-dependent scattering
cross-sections \citep{1993PhRvC..47..110J, 1995PhRvL..74..654H,
1998PhRvC..57...39I}.

Figure~1 indicates that the spin temperature of \he3 decouples from
the gas temperature at very high redshifts for regions at the average
cosmic density.  However, even for regions with an overdensity of
$\sim 10$, the spin temperature is only a few times the CMB
temperature $T_{cmb}$ through the EoR.  This factor alone suppresses
the signal from such regions by about $20\%$ to $50\%$.  Regions with
much higher overdensity decouple only at late times and emission from
such regions is not affected significantly at high redshifts.  For
example, at $z \simeq 3$ the low spin temperature leads to an
reduction in the signal by about $20\%$.  In general, the reduction in
signal due to a low spin temperature is negligible for regions with
electron number density $n_e \geq 10^{-2}~{\rm cm^{-3}}$ if the
\citet{1967ApJ...149...15G} estimate is reliable.

During reionization, brightness fluctuations on large scales (tens of
comoving Mpc or arc minutes on the sky) will be sourced by
inhomogeneities in the \he3 ionization fraction or in the kinetic
temperature of the gas, which is heated to $\sim 10^4$K in ionized
regions \citep{2008ApJ...689L..81T}.

The absorption signal could be searched for in the radio spectrum of
radio-loud quasars or gamma-ray burst afterglows
\citep{2005ApJ...619..684I}. Absorption is expected to be weak in most
situations, especially if the spin temperature is coupled to the
kinetic temperature.  The kinetic temperature in regions with
substantial fraction of \he3 is likely to be around $10^4$~K and hence
the optical depth due to absorption is extremely small in regions with
$T_s \simeq T_{\rm gas}$.  

The brightness temperature for emission in the hyperfine transition of
\he3 is given by,
\begin{eqnarray}
\delta T_b &\simeq& 18 \mu{\mathrm K} ~~ x_{He II} \left(1 -
 \frac{T_{cmb}}{T_s} \right)  \left[\frac{1+\delta}{10}\right]
 \left[\frac{1+z}{9}\right]^{1/2}  \nonumber \\ 
&& ~~~~~~~~~ \times ~~~~~~
\left(\frac{n_{^3He}/n_{H}}{1.1 \times 10^{-5}}
  \right)\left[\frac{H(z)}{(1+z) (dv_\parallel/dr_\parallel)}\right] 
\label{Tb}
\end{eqnarray}
Since the first ionization of helium and hydrogen are expected to be
contemporary \citep{2003ApJ...586..693W}, the power-spectrum of \he3
brightness fluctuations during the hydrogen EoR can be calculated in
analogy with the $21$~cm power-spectrum \citep{2008PhRvD..78j3511P}.  The
\he3 brightness temperature is smaller than that expected for the $21$~cm
transition of H~{\small I}.  However, the higher rest frame frequency
frequency compensates for the reduced signal, as the expected level of
foregrounds increases sharply at lower frequencies.  Thus, observing
this signal is not as difficult as might seem at a first glance.  Of
course, for a system to be seen in emission it must have $T_s >
T_{cmb}$.  Figure~1 may serve as a guide in gauging the significance
of the term $(1 - T_{cmb}/T_s)$.  We discuss specific sources in the
following section.

\section{Discussion}

Next we discuss prospects of observing systems the hyperfine
transition of \he3 in emission during and after the EoR.  The expected
signal is weak, and there is a further attenuation due to decoupling
of gas temperature and spin temperature in regions that are not very
overdense.  Thus, the most promising sources that may be detected and
studied are those that are sufficiently overdense.

Ly-$\alpha$ blobs, first discovered by \citet{2000ApJ...532..170S}, are
of particular interest. By now, many such blobs have been detected
\citep{2000ApJ...532..170S, 2006A&A...452L..23N, 2009ApJ...693.1579Y},
including one at a redshift $z=6.6$ \citep{2009ApJ...696.1164O}.
These blobs are hundreds of kpc across and contain gas at a high
overdensity.  The leading theoretical model for these blobs is
accretion of cold gas at temperatures around $10^{4}$~K along
filaments \citep{2009arXiv0902.2999D}.  Irrespective of the detailed
model, we can trivially deduce that gas in these blobs is at
temperatures around $10^4$~K as much hotter gas will not emit much in
Ly-$\alpha$, and that the overdensities in these blobs are fairly
high, certainly of the order of $10^2$ if not more.  This makes such
blobs a very strong source of radiation in the hyperfine transition of
\he3.  If we assume that $\delta \sim 10^2$--$10^3$ (a reasonable
assumption since we are dealing with gas inside collapsed haloes),
$x_{HeII} \sim 0.5$, and $[dv_\parallel/dr_\parallel]/[H(z)/(1+z)]
\sim 0.1-1$, the estimated \he3 signal is in the range of
$10$--$500$~$\mu$K.  Since a large number of Ly-$\alpha$ blobs are
known, a targeted study is possible.  Observations of blobs in the
hyperfine transition of \he3 can provide a direct handle on the the
physical properties of the sources.  The number density of Ly-$\alpha$
blobs is sufficiently high \citep{2009arXiv0902.2999D} so that blind
searches can also be undertaken in the redshifted radiation from the
hyperfine transition of {\he3}.

H~{\small II} regions around clusters of the earliest galaxies are
also an obvious source of this radiation.  For typical stellar
sources, $^3$He is expected to be ionized together with hydrogen since
its first ionization potential is only a factor of $1.8$ times larger
\citep{2003ApJ...586..693W}.  Observational constraints on
reionization models suggest that normal stars may suffice as a source
of ionizing radiation and enrichment of the IGM
\citep{2006ApJ...647..773D, 2009arXiv0902.0853B}.  We therefore expect
that tracing \he3 at redshifts $z>7$ is equivalent to tracing the
H~{\small II} regions.  A map of \he3 of a representative cosmic volume
during reionization would therefore largely appear as the contrast of
a $21$~cm map of the same volume.  Anti-correlating the two signals
would be helpful in strengthening the signal-to-noise ratio of both
maps.  In addition, a measurement of the \he3 signal can be used (in
conjunction with $21$~cm data) to determine the cosmic $^3$He
abundance as a test of big bang nucleosynthesis.  
Note that for sources at redshift $z \simeq 15$, the observed frequency is
close to $550$~MHz and is in the regime where the sky temperature is
subdominant to the system temperature for most existing and planned
instruments. 
At high redshifts, even regions with moderate overdensities can be seen in
emission. 

In case of sources with a harder spectrum, we expect an inner region
dominated by He~{\small III}.  The recombination time for He~{\small
III} is much smaller as compared to He~{\small II}.  This leads to
formation of fossil \he3 regions if the source of hard radiation
switches off for an extended period of time.  Quasars are the most
obvious source of hard UV radiation and these are known to have
intermittent periods of activity.  Therefore fossil \he3 regions at
high redshift are also a promising source of signal.

Several existing and upcoming instruments have the required frequency
coverage to attempt observations of the hyperfine transition in \he3.  We
summarize some of these here: 
\begin{itemize}
\item
The Giant Meterwave Radio Telescope
(GMRT)\footnote{http://gmrt.ncra.tifr.res.in} located near Pune,
India, has been in operation for over a decade and its performance is
being enhanced through upgrades.  The complete frequency range
$1-1.42$~GHz is available, allowing observations of the hyperfine
transition of \he3 between $5.1 \leq z \leq 7.6$, at least in
principle.
The best sensitivity achieved so far is $30$~$\mu$Jy\footnote{See page
  9 of
  http://www.ncra.tifr.res.in/$\sim$gtac/GMRT\_specs\_status\_Dec08.pdf}.
This is close to two orders of magnitudes larger than the sensitivity
required to observe Ly-$\alpha$ blobs, hence it will be difficult to use the
GMRT for observing Ly-$\alpha$ blobs at high redshifts even with long
integrations on a single source, unless upcoming upgrades can reduce the noise
level by a significant amount.
\item
The Expanded Very Large Array
(EVLA)\footnote{http://www.aoc.nrao.edu/evla/}, is an upgraded version
of the very large array (VLA).  This facility will enter a testing
phase with science projects in $2010$ and is likely to be available
for all users by $2013$.  EVLA will have full coverage from $1$~GHz to
$50$~GHz and hence can be used to probe \he3 redshifts up to $z=7.6$.
The low frequency system is to be installed last.  The overall sensitivity
is slightly less than that for ASKAP.
\item
The Australian Square Kilometer Array Pathfinder
(ASKAP)\footnote{http://www.atnf.csiro.au/projects/askap/} will
comprise an array of $36$ antennas each $12$~m in diameter, capable of
high dynamic range imaging and using wide-field-of-view phased array
feeds.  This observatory is expected to become operational by $2013$.
Frequencies between $700$~MHz and $1.8$~GHz will be accessible using
ASKAP indicating that it may be possible to observe up to $z \sim 11.4$.  The
stated sensitivity will be insufficient for observing Ly-$\alpha$ 
blobs with around $10^3$ hours of integration time.  However, the field of
view for this instrument is large and it may be possible to run a piggyback
survey where co-adding the signal from many known Ly-$\alpha$ blobs may be
used to enhance the signal-to-noise ratio. 
\item
The Square Kilometer Array
(SKA)\footnote{http://www.skatelescope.org/} is expected to have an
effective collecting area that is nearly two orders of magnitudes
larger than SKA pathfinders like the ASKAP.  This alone ensures that
observations of Ly-$\alpha$ blobs will be possible with the SKA with
an integration of just above a hundred hours.
\end{itemize}

Finally we comment on the issue of foregrounds.  The sky temperature
$T_{sky}$ is a very strong function of frequency at low frequencies,
scaling as $\nu^{-2.7}$ at low frequencies ($\la 1$~GHz).  Given the
difference in rest frame frequencies of the hyperfine transition of
neutral hydrogen and \he3, the expected difference in the sky
temperature for observing the same redshift differs by a factor of
almost $200$. This nearly compensates for the difference in the
expected signals for the two transitions at the highest redshifts of
interest for the epoch of reionization studies.  At lower redshifts,
$T_{\rm sys} > T_{\rm sky}$ for most upcoming arrays and the
instrumental noise is the dominant factor.  Furthermore, the higher
frequency implies than the Sun and ionosphere are not very serious
problems and day time observations may also be possible, thereby
increasing the possible integration time on one source.

For redshifts beyond $5.1$, H~{\small I} at low redshifts is a foreground
contaminant for the hyperfine transition of {\he3}.  
This is not likely to be a cause for concern for targeted studies but is an
issue for statistical detection. 
One may either choose {\it clean} lines of sight, or use differences in the
angular and redshift space distribution of foregrounds as compared to the
signal. 

Recombination lines of hydrogen can be a source of confusion,
particularly if the number densities of electrons are more than a few
per cm$^{-3}$.  Observations of H~{\small II} regions
\citep{2006ApJ...640..360B} find several lines of this type near the
\he3 hyperfine transition.  The impact of such lines on cosmological
observations has been found to be small by \citet{2003MNRAS.346..871O}.

\section{Summary}

In this paper we have considered the use of the hyperfine transition
of \he3 for studies of the IGM, with an emphasis on the possibility of
detecting it in emission.  The frequency of the \he3 transition is
$\sim 6$ times larger while its radiative decay rate is almost three
orders of magnitude higher than those of the $21$~cm transition of
hydrogen.  These parameters lead to a weaker coupling between the
kinetic temperature of the IGM and the \he3 spin temperature.
However, \he3 naturally exists in regions where the gas temperature is
relatively high ($\sim 10^4$K) and even this weak coupling can ensure $T_s \gg
T_{cmb}$ in moderately overdense regions.

Although the abundance of $^3$He is around $10^{-5}$ times that of
hydrogen, its hyperfine signal is boosted by its higher decay rate.
Additional gain in the signal-to-noise ratio originates from the lower
sky temperature at the frequency of the \he3 transition as compared
with the sky temperature for the $21$~cm radiation at the same
redshift.  Thus, observing the \he3 signal may not be much more
challenging than observing the redshifted $21$~cm signal.

The higher rest frame frequency of the hyperfine transition of \he3 also
allows one to probe physical scales that are six times smaller, as compared to 
those that can be accessed with the $21$~cm radiation using the same
instrument for a region at the same redshift.

We find that the extended Ly-$\alpha$ blobs seen at high redshifts are
the most promising source in this transition and that \he3
observations may shed new light on the physical conditions in these
regions.  A number of Ly-$\alpha$ blobs are already known and it is
possible to conduct a targeted study.  The expected signal of
$10$--$200\mu$K should be within the reach of upcoming instruments.
In the longer term, observations of the hyperfine transition in \he3
may be used to probe cosmological H~{\small II} regions during the
EoR.

\section*{Acknowledgments}

This work was supported in part by Harvard University funds and NASA
grant NNX08AL43G.  This research has made use of NASA's Astrophysics
Data System.  We acknowledge the use of the Legacy Archive for
Microwave Background Data Analysis (LAMBDA).  Support for LAMBDA is
provided by the NASA Office of Space Science.

While this project was nearing completion, we learned of a similar
effort on \he3 hyperfine emission by M. McQuinn, \& E. R. Switzer
(2009, submitted to Phys. Rev. D).  We refer the reader there for
complementary discussion.

\label{lastpage}

\end{document}